\magnification=1200
\overfullrule=0pt
\baselineskip=20pt
\parskip=0pt
\def\dag{\dagger}
\def\del{\partial}

\def\a{\alpha}             
\def\b{\beta}              
\def\g{\gamma}        
\def\d{\delta}        
\def\e{{\rm e}}

\def\x{\xi}              
                  
\def\p{\pi}              
\def\r{\rho}               
\def\s{\sigma}

\def\y{\psi}               

\def\yd{\y^{\dag}}

\def\yh{\hat{\y}}
\def\yhd{\hat{\y}^{\dag}}

\def\w{\omega}     
   
\def\br{\langle}
\def\ke{\rangle}
\def\ve{\vert}

\def\to{\rightarrow}

\def\zbar{\bar{z}}
\def\to{\rightarrow}
\def\eqref#1{(\ref{#1})}
\def\zintm{d^2 z \, \e^{-|z|^2}}

\def\tr{{\rm tr}}

{\settabs 5 \columns
\+&&&&CCNY-HEP-96/8\cr
\+&&&&LPTENS-96/44\cr
\+&&&&July 1996\cr}
\bigskip
\centerline{\bf COLLECTIVE VARIABLES OF FERMIONS AND BOSONIZATION}
\bigskip\bigskip
\centerline{ B. Sakita$^*$}
\bigskip
 
\centerline{ Physics Department, City College of the 
City University of New York}
\centerline{ New York, NY 10031}
\bigskip
\centerline{\bf Abstract}
\bigskip
We first present a general method for extracting  collective variables out of
non-relativistic fermions by extending the gauge theory of collective
coordinates to fermionic systems. We then apply the method to a system of 
non-interacting
flavored fermions confined in a one-dimensional flavor-independent potential.
In the limit of a large number of particles
we obtain a Lagrangian with the Wess-Zumino-Witten term, which is the 
well-known 
Lagrangian describing
the non-Abelian bosonization of chiral fermions on a circle.
The result is universal and does not depend on the details of the confining potential.

\vfill

\noindent{$^*$E-mail address: sakita@scisun.sci.ccny.cuny.edu.}
\vfill\eject
\noindent{\bf 1. Introduction}

An interesting approach to the bosonization of non-relativistic
fermions is through the application of collective field theory to the 
large $N$ limit of the
$D=1$ matrix model [1], since the model is known to be equivalent to 
a system of $N$ non-interacting non-relativistic fermions in one space 
dimension [2]. Unfortunately, however, this approach is impractical,
because not only does the model represent a fermionic 
system strictly in one space dimension but 
it also fails to include spin and interactions.
In this paper we propose an alternative approach,
which does not restrict extensions and generalizations while preserving some 
good features of collective field theory.

Soon after the method of collective coordinates was proposed
to resolve the difficulty associated with the zero-modes in soliton quantization [3],
Hosoya and Kikkawa presented a gauge theoretical formulation of
collective coordinates [4]. It was then noticed that the method 
was general enough to handle even the non-zero modes [5]. 
Since the term {\it collective
coordinates} is used exclusively for zero modes now-a-days, in this paper
we use the term {\it collective variables} for non-zero modes in order to avoid
possible confusion. In this article,
we extend the gauge theory of collective coordinates
to a system of non-relativistic
fermions in order to extract the collective variables of fermions. As we shall
describe in section 2, we extract 
the collective variables to such an extent that all the dynamics is
expressed in terms of them and that, 
at the end, all the fermionic degrees of freedom are frozen. 
In this sense the spirit
of the present paper is the same as that of collective field theory [1].
The Lagrangian we obtain in section 2 is similar, if not identical, to that
written down previously by a number of authors [6].

In order to stress the usefulness as well as the validity of the theory, 
in section 3 we apply
the method to a system of $N$ non-interacting flavored fermions, which
are confined in a flavor-independent one-space-dimensional potential. 
We consider this model as a non-Abelian extension of the matrix model.
In the large $N$ limit
we derive a
Lagrangian, which involves the Wess-Zumino-Witten term [7]. This is the
well-known Lagrangian resulting from the non-Abelian bosonization of 
chiral fermions on a circle [8].

\bigskip
\noindent {\bf 2. Collective Variables of Fermions}

In order to extract the collective degrees of freedom of
fermions, we extend the gauge theory of collective coordinates to 
a system of fermions.

Let us denote by $\a \ (\a =
1, 2,\cdots , K)$ all the attributes of the fermion as well as the position. 
From now on we call $\a$ simply a position. 
Let $\yh _\a $ and $\yhd _\a $ be the annihilation
and creation operator respectively of the fermion at $\a$.
We assume that the fermion number is conserved and the Hamiltonian 
of the system is expressed as
$$
\hat{ H }\equiv H\{\yhd ,\yh\}=\sum_{\a.\b}\yhd _\a  h^{(1)\b}_{\a}\yh _\b  +
\sum_{\a_1 ,\a_2 ,\b_1 ,\b_2}\yhd _{\a_1 }
\yhd _{\a_2 } h^{(2)\b_1 \b_2}_{\a_1 \a_2}
\yh _{\b_1 }\yh _{\b_2 }+\cdots \eqno (2.1)
$$
The classical Lagrangian from which the Hamiltonian (2.1)
is derived upon the canonical quantization is given by
$$
L=i\sum _{\a}\yd_\a {d\over{dt}}{\y}_\a - H\{\yd ,\y\} \ .\eqno (2.2)
$$

Next we consider a Lagrangian given by
$$
L=i\sum_{\a ,\b , \g}\yd_\a u^{\dag}_{\a\b}
{d\over{dt}}\big( u_{\b\g}{\y}_\g \big) - H\{\yd u^{\dag}, u\y\}\ ,\eqno (2.3)
$$
where $u$ is a time dependent $K\times K$ unitary matrix. The Lagrangian (2.3)
is merely the Lagrangian (2.2) in which $\y$ is replaced by $u\y$. But,
in (2.3) we consider both $u$ and $\y$ as dynamical variables.
It is obviously invariant under the unitary
transformation
$$
\y \to v\y ,\ \ \ \ \ u\to uv^{\dag}\ , \eqno (2.4)
$$
where $v$ is a time dependent $K\times K$ unitary matrix.
The theory is thus an $SU(K)$ gauge theory.

We express $u$ in terms of $A$ as $u(A)\equiv\exp iA$, where $A$ is a 
$K\times K$ hermitian matrix: $(A)_{\a\b}\equiv\sum_{a=0}
^{K^2 -1} (t_a )_{\a\b}A_a$.  We use a set of normalized $t_a\, (a=0,1,2\cdots 
K^{2 }-1 )$,
a fundamental representation of $U(K)$ generators, such that 
$\sum_a (t_a )_{\a\b}(t_a )_{\g\d} =\d_{\a\d}\d_{\b\g}$.
Since the theory is a gauge theory, we obtain a set of first class
constraints:
$$
P_a -\sum_b {\cal N}_{a}^b (A) \yd t_b\y \approx 0  \ \ \ (a = 0, 1, \cdots ,
K^2 -1 ) \eqno (2.5)
$$
where $P_a$ is the momentum conjugate to $A_a$ and ${\cal N}(A)$ is a $K^2\times K^2$
matrix defined by $-i{\del\over{\del A_a}}\big( u(A)\big)_{\a\b}
=\sum_b {\cal N}_a ^b (A)\big( u(A) t_b \big)_{\a\b}$.
If we choose the gauge $A_a =0$ we obtain the original fermionic theory. If
we choose a gauge that freezes the fermion degrees of freedom, we obtain
a theory described by $A$, namely a bosonization. 

In the Hamiltonian formalism we impose the Gauss law constrains
on the physical states. This amounts to saying that the wave functions are
functions of gauge invariants, in the present case $u(A)\y$. 
Since $\y$ is a fundamental representation of $SU(K)$,
one can rotate $\y$
into a configuration in which only the lower $N$ components are non-zero. 
After the rotation one lets the transformation matrix be 
absorbed into $u(A)$ by redefining the variable $A$. This procedure 
is nothing but a gauge fixing procedure. Since the $N$ particle state
of $N$ component fermion is a single state, the fermion
degrees of freedom become trivial.

The gauge fixing procedure described above corresponds to the gauge
$$
\y_\a =0,\ \ \ \ \y^\dag _\a =0\ \ \ \ \ (\a >N ) .\eqno (2.6)
$$
For a system of $N$ particles, all the available positions
of fermions in this gauge are filled and the excitations of the system are solely
described by $A$. In effect we can set
$$
\yd_\a \y_\b = (\r_0 )_{\a\b},\ \ \ \  \yd_\a\yd_\b\y_\g\y_\d =
(\r_0 )_{\a\d}(\r_0 )_{\b\g}-(\r_0 )_{\a\g}(\r_0 )_{\b\d},\ \cdots
\eqno (2.7)
$$
where $\r_0$ is a diagonal matrix where the lower $N$ elements are $1$ 
and others are $0$.
If we use (2.7) in the Lagrangian (2.3) we obtain
$$
L=\tr\big(\r_0 u^\dag (i\del_t -h^{(1)})u\big)-\sum_{\a\b\g\d}
(u\r_0 u^\dag )_{\a\b}(u\r_0 u^\dag )_{\g\d}\big(h_{\b\d}^{(2)\a\g}
-h_{\b\d}^{(2)\g\a}\big) -\cdots \ .\eqno (2.8)
$$
Lagrangians of this kind have already been proposed in [6] for a
description of a fermionic system in terms of bosonic variables.

The gauge condition (2.6) does not fix the gauge completely, since (2.6) and (2.7)
are 
invariant under $SU(N)\times SU(K-N)$ transformation. 
We observe that the change in the Lagrangian
(2.8) under $u(A)\to u(A)v$ is a total time derivative if $v$ is a matrix of
$SU(N)\times SU(K-N)$ transformation.
After the gauge fixing the wave function must be a 
singlet under $SU(N)$. Since in the bosonized theory this condition
is a consequence of the residual $SU(N)$ gauge symmetry,
we must keep this symmetry intact.

\bigskip
\noindent {\bf 3. Non-Abelian Bosonization of Fermions in 
One Space Dimension}

In this section we consider a system of non-interacting fermions confined in
a one-space-dimensional potential.
We assume that 
the system has an internal symmetry, which we specify by flavored fermions
and a flavor-independent Hamiltonian:
$$
h^{(1)}=\hat h ={1\over 2}{\hat p^2}+ v(\hat x )\ .\eqno (3.1)
$$
We use the Lagrangian (2.8) to describe a
non-Abelian bosonization of this system by appropriately extending the range
of $K$ and $N$. 

We extend
the index $\a$ of $K$ to a pair of $x$, a continuous space-coordinate,
and $i$, a flavor-index.
The hermitian matrix $A$ is then replaced by a bi-local
matrix field $A_{ij}(x,y)$, which we consider as a matrix element of an operator
$\hat A_{ij}$:
$$
A_{ij}(x,y)=\br x\ve \hat {A}_{ij} \ve y\ke \ .\eqno (3.2)
$$
We express $\hat A_{ij}$
by using the operators of a harmonic oscillator:
$$
\hat {A}_{ij}\equiv {A}_{ij} (\hat{a},\hat{a}^\dag )
=\ddag {A}_{ij}(z,\zbar ){\bigm | }_{{z=\hat{a}\ }\atop{\zbar =
\hat{a}^\dag}}
\ddag =\int\zintm\ve z \ke {A}_{ij}
 ( z,\zbar )\br z\ve\ , \eqno (3.3)
$$
where $\hat{a}^\dag$ and $\hat a$ are the standard 
creation and annihilation operators and   
$\ddag\ \ \ddag$ is the anti-normal-order symbol,
i.e. all the creation operators 
stand to the right of the annihilation operators.
The last expression in (3.3) is in the coherent state representation [9].
From now on we simply call ${A}( z,\zbar )$ 
the coherent state representation of $\hat A$ and furthermore we suppress 
the flavor indices. The coherent state representation
of a product of two operators is given by the star product [9]:
$$
\eqalign{
&\hat A \hat B =\int\zintm\ve z \ke A( z,\zbar ) * B ( z,\zbar )\br z\ve \ ,\cr
&A( z,\zbar ) * B ( z,\zbar )
=
\sum_{n=0}^{\infty} {{(-)^n}\over {n!}}\big(\del_{\zbar}^n A( z,\zbar )\big)
\big(
\del_z ^n  B ( z,\zbar )\big) \ .\cr} \eqno (3.4)
$$
The star products are associative.

In order to specify the matrix
$\r_0$ we use the number representation of the harmonic oscillator:
$$
\hat\r_0\ =\ \sum_{n=0}^{N-1} \ve n\ke\br n\ve\ , \eqno (3.5)
$$
which corresponds to the state where the lower $N$ levels are completely filled.
It is a flavor singlet state.
We define $\r_0$ by
$$
\r_0\equiv\r_0 (|z|^2 ) = \br z | \hat\r_0 |z\ke {\rm e}^{-|z|^2}=
\sum_{n=0}^{N-1} {{|z|^{2n}}\over n!}{\rm e}^{-|z|^2}\approx \theta ( N
-|z|^2 ) \eqno (3.6)
$$
The last expression is an approximation valid in the large $N$ limit [10].

\bigskip
\noindent {\bf Calculation of
$i\tr\big( \hat \r_0 u^\dag (\hat A ){d\over{dt}} u(\hat A )\big)$ for large ${\bf N}$}

We first calculate $\tr\bigg(\hat \r_0 \hat{A}^n {d\over{dt}}\big(\hat {A}^{n'}
\big)\bigg)$ for large $N$.
Using the coherent state representation of $A$ we express it in an integral form:
$$
\tr\bigg(\hat \r_0 \hat{A}^n {d\over{dt}}\big(\hat {A}^{n'}\big)\bigg)
=
\int d^2 z \,\r_0\,\tr\big( A^{*n} * \del_t A^{*n'}\big) \ ,
$$
where $A^{*n}$ is the star product of $n$ $A$'s and the trace 
in the right hand side of equation is for flavor indices. 
In this integral if we change variables from $z$ to $\s={1\over{\sqrt N}}z$ and 
eliminate the $N$ dependence in $\r_0$, each derivative in the star product
brings a factor ${1\over{\sqrt N}}$. Assuming that $A$ and its $\s$ derivatives
are of order $1$, we therefore keep only the terms which
contain two powers of $z,\zbar$ derivative or less. This assumption is reasonable
because in the large $N$ limit $A(\vec\s)$ can be considered as a field of quantum
fluctuations (see the discussion preceding eq. (3.12)). 
We obtain
$$\eqalign{
\tr\big(\hat \r_0 &\hat{A}^n {d\over{dt}}(\hat {A}^{n'})\big)\cr
&\approx\int d^2 z \,\r_0\, \tr\big( A^n  
{\del_t} A^{n'} -(\del_{\zbar} A^n )
{\del_t \del_z}A^{n'}
- (A^n )_{,\zbar ,z}\del_t
A^{n'}- A^n \del_t 
( A^{n'})_{,\zbar ,z}\big) ,\cr}
$$
where
$$
(A^n )_{,\zbar ,z}=\sum_{p=0}^{n-1}
A^p ( \del_{\zbar} A)\del_z (
A^{n-p-1} ) \ .
$$
Substituting the above into the power series expansion of
$i\tr\big( \hat \r_0 u^\dag (\hat A ){d\over{dt}} u(\hat A )\big)$, we obtain
$$\eqalign{
\tr\big(\hat \r_0\hat u^\dag &{d\over{dt}}\hat u\big)\cr
\approx&\int d^2 z \ \r_0 \ \tr\big( g^{-1}  
{\del_t} g -
(\del_{\zbar}g^{-1} )
{\del_t \del_z}g
-( g_{,\zbar ,z})^\dag
\del_t g-g^{-1}  
\del_t g_{,\zbar ,z}\big)\cr}\eqno (3.7)
$$
where $g $ and  $g_{,\zbar ,z}$ are given by
$$\eqalign{g&=\sum_{n=0}^{\infty}{{(-i)^n}\over{n!}}
A^n =\exp(iA)\cr
g_{,\zbar ,z} &=\sum_{n=0}^{\infty}{{(-i)^n}\over{n!}}
\sum_{p=1}^{n-1} A^p (\del_{\zbar}
A )\del_z (A^{n-p-1} )\cr
&=ig\int_0 ^1 d\a {\rm e}^{-i\a A}(\del_{\zbar} A)\del_z {\rm e}^{i\a A}\cr
&=-g\int_0 ^1 d\a {\rm e}^{-i\a A}(\del_{\zbar} A){\rm e}^{i\a A}
\int_0 ^\a d\b {\rm e}^{-i\b A}(\del_{\zbar} A) {\rm e}^{i\b A}\ .\cr}\eqno (3.8)
$$

The first term of (3.7) is a total time derivative, since
$$
 dg =\sum_{n=0}^{\infty}{{(-i)^n}\over{n!}}
\sum_{p=1}^{n-1} A^p (d
A )A^{n-p-1} =i\, g\, \int_0 ^1 d\a {\rm e}^{-i\a A}dA{\rm e}^{i\a A}\ . 
$$
Using this and the last expression in (3.8),
one can prove that
$$
g^{-1}  g_{,\zbar ,z} +\big( g^{-1}  g_{,\zbar ,z}\big)^{\dag}
=(g^{-1}\del_{\zbar} g)(g^{-1}\del_z g)\ .\eqno (3.9)
$$ 
By a repeated use of
$dg^{-1}=-g^{-1}dgg^{-1}$ and (3.9) one can also prove that
$$\eqalign{
\tr\big( &g^{-1} \del_t g_{,\zbar ,z} + (g_{,\zbar ,z} )^\dag
\del_t g + \del_{\zbar} g^{-1} \del_z\del_t g \big)\cr
&={1\over 2}\big[\del_{\zbar}\tr ((g^{-1} \del_z g)( g^{-1} \del_t g) \big)-
\del_z \tr ((g^{-1} \del_{\zbar} g) (g^{-1} \del_t g )\big)\big]\cr
&\ \ \ \ +{1\over 2}\tr\big([(g^{-1} \del_{\zbar} g) , (g^{-1} \del_z g )] (g^{-1}
\del_t g)\big)+
{\rm total\  time\  derivative}\ .\cr}\eqno (3.10)
$$
Substituting this into the integrand of (3.7) and integrating it by part
we finally obtain
$$
\eqalign{
i\tr\big( \hat \r_0 u^\dag (\hat A ){d\over{dt}} u(\hat A )\big)
=-{1\over {4\p}}&\oint d\theta \tr\big((g^{-1}\del_\theta g)(g^{-1}\del_t g)\big)\cr
&\ \ \ -{1\over{4\p}}\int_0^1 dr\oint d\theta\tr\big([(g^{-1}\del_\theta g),(g^{-1}\del_r
g)](g^{-1}\del_t g)\big)\ .\cr} \eqno (3.11)
$$
The last term is the Wess-Zumino-Witten action [7].

\bigskip
\noindent{\bf Calculation of $\tr\big(\hat\r_0 \hat u^\dag \hat h \hat u\big)$
for large N}

We notice that the result (3.11) does not change even if  
$\hat u$ is replaced by $u(\hat \xi ) u(\hat A )$ provided $\xi$ is time-independent.
We make this modification and
determine $\xi$ by the condition that the energy of the system is at minimum
when $A=0$. This is the same procedure that we normally follow in the
semi-classical expansion of ordinary
scalar field theories, namely $\xi$ corresponds
to a shift to the minimum of the potential
and $A$ corresponds to the quantum fluctuations about it.
The shift can be as large as of order $N$, but the fluctuations are of order $1$.
We assume $\hat h$ is given by (3.1) and is flavor-independent.
\def\hub{\underline{h}}
We define $\hat{\hub}$ by
$$
\hat{\hub} = u^\dag (\hat\xi ) \hat h u(\hat\xi ).\eqno (3.12)
$$
We obtain
$$
\tr\big(\hat\r_0 \hat u^\dag \hat h \hat u\big)=\int d^2 z\,\r_0\,\sum
_{n=0}^\infty {{(-i)^n}\over{n!}}\tr\big( [A,*[A,*[\cdots [A,*\hub ]\cdots ]]]\big)\ ,
\eqno (3.13)
$$
where $\hub (z,\zbar)$ is the coherent state representation of
$\hat{\hub}$ and is given by
$$
\hub (z,\zbar )=
\sum
_{n=0}^\infty {{(-i)^n}\over{n!}} [\xi ,*[\xi ,*[\cdots [\xi ,
*h ]\cdot ]]]\ .\eqno (3.14)
$$

We first calculate the $n=0$ term in the expansion (3.13), namely 
$$
\int d^2 z\,\r_0\,
\hub (z, \zbar )\ .\eqno (3.15)
$$ 
Since both $h$ and $\r_0$ are flavor-independent, it is sufficient
to consider flavor-independent $\x$. Then the commutator $-i[\ \ ,*\ \ ]$ in (3.14)
is a Moyal bracket.
We calculate the integral in the large $N$ limit. For this purpose 
we change variables from $z$ to $\s$ before taking the large $N$ limit:
$$
z=\sqrt{N}\s,\ \ \ \x=N\tilde\x (\vec \s ),\ \ \  h(z,\zbar )=N\tilde h (\vec\s ),
\ \ \ 
\hub (z,\zbar )=N\tilde{\hub}(\vec\s )\ .\eqno (3.16)
$$
This large $N$ limiting procedure is the same as the $W_\infty\to w_\infty$
contraction procedure described in [9]. We obtain
$$
\tilde{\hub}(\vec\s )=\sum_{n=0}^\infty {1\over{n!}}
\{\tilde\xi , \{\tilde\x ,\{\cdots\{\tilde\x ,\tilde h\}\cdot\}\}\}
=\exp\big(\epsilon_{ij}\del_i\tilde\x \del_j\big)\tilde h (\vec\s )\ ,\eqno (3.17)
$$
where $\{\ \ ,\ \ \}$ is a Poisson bracket.
The transformation $\tilde h\ \to \ \tilde{\hub}$ is therefore 
a canonical transformation.

We describe this canonical transformation by using a familiar notation
that one uses
in classical mechanics. Let $q,p\ \to\ Q, P$  be a canonical transformation
such that $\tilde h(q,p)={1\over 2}p^2 +U(q)\ \to \ \ \tilde{\hub}
(Q, P)=\tilde{h}(q(Q,P),p(Q,P))$,
where $ U(x)={1\over N} v(\sqrt{N} x)$.  
With this notation
the integral (3.15) is expressed as
$$
\int_{Q^2 + P^2 < 2} {{dQdP}\over{2\p}}\tilde{\hub} (Q,P) =\int_{V}{{dqdp}\over{2\p}}
\tilde{h}(q,p)\ .\eqno (3.18)
$$
The boundary of $V$ depends on the transformation, namely on $\x$, but the
area of $V$ must be $2\p$. The boundary, which makes the integral 
of the Hamiltonian minimum
for a given area, is a curve of constant energy:
$$
{1\over 2}p^2 +U(x)=e_0 ,\ \ \ \int_{x_1}^{x_2}dx \sqrt{2(e_0 -U(x))}=\p \ ,\ \ \
U(x)={1\over N} v(\sqrt{N} x)\ .\eqno (3.19)
$$
At this energy, this analogous classical mechanical system undergoes a periodic
motion with a half period $T$ and an angular frequency $\omega$ given by
$$
T=\int_{x_1}^{x_2}{{dx}\over{ \sqrt{2(e_0 -U(x))}}} ,\ \ \ \omega ={\p\over T}\ .
\eqno (3.20)
$$
In the $Q,P$ coordinate this motion should be described by Hamiltonian $\tilde{\hub}
(Q,P)$.
Since the trajectory of the motion in phase space is
given by $Q^2 + P^2 =2$ and since $\omega$ is invariant under the canonical 
transformation,
$\tilde{\hub}$
near the boundary must be given by 
$\omega {1\over 2}(Q^2 + P^2 ) + {\rm const. }$.
Thus we obtain 
$$
\hub (z,\zbar ) = \omega |z|^2 + \ {\rm const.} \ \ \ \ {\rm for }
\ \ \  |z|^2\sim
N\ \  \eqno (3.21)
$$
It is interesting to notice that (3.21) is quite universal since
only the value of $\w$ depends on the potential.  

Now we are ready to calculate (3.13). We first notice that the leading term of
an integral of a trace of star commutator is in general given by
$$
\int d^2 z \r_0 \, \tr\big( [A, *C]\big)\ \approx\ -i\oint{{d\theta}\over{2\p}}
\tr\big( (\del_\theta A) C\big)\ ,\eqno (3.22)
$$
and also the leading term of 
$$
[A,*[A,*\cdots [A,*[A,*\hub ]]\cdot ]]\approx - i\omega [A,[A,\cdots [A,\del_\theta A]
\cdot ]] \ .\eqno (3.23)
$$
Using these we finally obtain
$$\eqalign{
\tr\big(\hat\r_0\hat u ^\dag \hat h \hat u\big)&= -{\omega\over{2\p}}\oint d\theta
\sum_{n=2}^\infty {{(-i)^n}\over{n!}}\tr\big((\del_\theta A)
[A,[A,\cdots [A,\del_\theta A]\cdot ]]\big)\cr
&=-{\omega\over{4\p}}\oint d\theta\tr\big((g^{-1}\del_\theta g)^2 \big)\ .\cr}
\eqno (3.24)
$$
The last step in (3.24) is confirmed by a straightforward comparison between
the series expansions.

The final result of the non-Abelian bosonization is thus
$$\eqalign{
L=-{1\over{4\p}}\oint d\theta&\tr\big((g^{-1}\del_\theta g)(g^{-1}(\del_t 
-\omega\del_\theta )g)\big)\cr
&-{1\over{4\p}}\int_0^1 dr\oint d\theta\tr\big([(g^{-1}\del_\theta g),(g^{-1}\del_r
g)](g^{-1}\del_t g)\big)\ .\cr}\eqno (3.25)
$$

\bigskip
\noindent {\bf 4. Discussions}

In this paper we have presented a bosonized theory of fermions, which 
becomes semi-classical for a system
of large number of fermions and admits of the $1/N$ expansion as
a semi-classical expansion. 

In section 2 we have derived the bosonized
Lagrangian (2.8) by extending the gauge theory of collective coordinates
to fermion systems. Although the same Lagrangian could be obtained by other means [6],
our derivation illuminates the role of the residual gauge symmetry 
hidden in the Lagrangian.
In our opinion, keeping this gauge symmetry intact is crucial for a successful
bosonization.

In section 3 we have discussed a bosonization of non-interacting
flavored fermions confined in a one-space-dimensional potential. 
The result we obtain is the Lagrangian describing the
non-Abelian bosonization of chiral fermions on a circle [8]. 
In this section
we have  extensively used the technique of the coherent state representation 
of the $W_\infty$
algebra that we have developed in a series of papers [9-12], without which we
could not have obtained the result.
The result is, however, an expected one based on the following arguments.

We first observe that for a large class of potentials the energy levels
near the $N$-th level, for $N$ large, 
are equally spaced.
This observation is equivalent to the statement about the universal form 
of the harmonic oscillator Hamiltonian 
obtained in (3.21). A simple proof goes
as follows. Let us denote by $E_0$ the energy of
$N$-th level and by $E_0 +\omega_n$ the energy of $(N+n)$-th level.
We use Bohr and Sommerfeld's quanization rule for both levels and expand the
equation for $(N+n)$-th level assuming $E_0\gg\omega_n$ for $N\gg n$. We obtain
$\w_n =n\w$, where $\w$ is given by (3.20). (QED).  
The linearity in energy dispersion is quite universal 
and it does not depend on the details
of the potential.
A sufficient condition is that the potential $U$
defined by (3.19) is smooth at the turning points.
Once the energy dispersion is
linear and bottomless, we know from Tomonaga's work [13] that 
the excitations of fermions are described by chiral bosons in one
space dimension, compact for discrete energy and non-compact for continuous.
The model in section 3 is an extension including an internal symmetry,
namely a non-Abelian generalization.

\bigskip
\noindent{\bf Acknowledgments}

I am thankful for the support given by LPTENS, where a part 
of the work was done.
I acknowledge valuable correspondences with S. Iso.
This work is supported by the NSF grant (PHY-9420615) and the Professional 
Staff Board 
Congress of Higher Education of the City University of New York under Grant
6-65399. 
\bigskip
\noindent{\bf References}
\item {[1]}
A. Jevicki and B. Sakita, {\it Nucl. Phys.} {\bf B165} (1980) 511. See also
\item {}
B. Sakita, {\sl "Quantum Theory of Many-Variable Systems and Fields"},
World Scientific, Singapore (1985).
\item {[2]}
E. Brezin, C. Itzykson, G. Parisi and J. B. Zuber, {\it Comm. Math. Phys.} {\bf 59 }
(1978) 35
\item {[3]}  
J.L. Gervais and B. Sakita, {\it Phys. Rev.} {\bf D11} (1975) 2943.
\item {[4]}
A. Hosoya and K. Kikkawa, {\it Nucl. Phys.} {\bf B101} (1975) 271.
\item {[5]}
J.L. Gervais, A. Jevicki and B. Sakita, {\it Phys. Reports} {\bf 230} (1976) 281.
\item {[6]}
J. P. Blaizot and H. Orland, {\it Phys. Rev.} {\bf C24} (1981) 1740.
\item {}
H. Kuratsuji and T. Suzuki, {\it Prog. Theor. Phys. Suppl. }{\bf 75} \& {\bf 76}
(1983) 209.
\item {}
A. Dhar, G. Mandal and S. R. Wadia, {\it Mod. Phys. Lett. }{\bf A8} (1993) 3557.
\item{[7]}
E. Witten, {\it Comm. Math. Phys.} {\bf 92} (1984) 455.
\item {[8]}  
J. Sonnenschein, { \it Nucl Phys.} {\bf B309} (1988) 752.
\item {}
M. Stone, {\it Phys. Rev. Lett.} {\bf 63} (1989) 731;
{\it Nucl. Phys. }{\bf B327} (1989) 399.
\item {[9]}
An. Kavalov and B. Sakita, CCNY-HEP-96/4, { hep-th/9603024}.
\item {[10]}
B. Sakita, {\it Phys. Lett.} {\bf B315} (1993) 124.
\item {[11]} 
S. Iso, D. Karabali and B. Sakita, {\it Nucl. Phys.} {\bf B 388} (1992) 700;
 {\it Phys. Lett.} {\bf B296} (1992) 143.
\item {[12]}
R. Ray and B. Sakita, {\it Ann. Phys.}{\bf 230} (1994) 131.
\item {[13]}
S. Tomonaga, {\it Prog. Theor. Phys. } {\bf 5} (1950) 544.
\item {}
M. Stone, {\sl "Bosonization"}, World Scienticic, Singapore (1994).

\end